\documentstyle[preprint,prd,aps]{revtex}
\input epsf

\global\arraycolsep=2pt
\makeatletter
\def\fmslash{\@ifnextchar[{\fmsl@sh}{\fmsl@sh[0mu]}}
\def\fmsl@sh[#1]#2{%
  \mathchoice
    {\@fmsl@sh\displaystyle{#1}{#2}}%
    {\@fmsl@sh\textstyle{#1}{#2}}%
    {\@fmsl@sh\scriptstyle{#1}{#2}}%
    {\@fmsl@sh\scriptscriptstyle{#1}{#2}}}
\def\@fmsl@sh#1#2#3{\m@th\ooalign{$\hfil#1\mkern#2/\hfil$\crcr$#1#3$}}
\makeatother

\begin{document}
\draft\pagenumbering{roma}
%%%%%%%%%%%%%%%%%%%%%%%%%%%%%%%%%%%%%%%%%%%%%%%%%%%%%%%%%%%%%%%%%%%%%%%%%%%%%
\title{Improved analysis for the baryon masses to order $\Lambda_{QCD}/m_Q$ \\from
 QCD sum rules}
\author{ Dao-Wei Wang$^a$, Ming-Qiu Huang$^{a,b}$ and Cheng-Zu Li$^a$}
\address{Department of Applied Physics, Nat'l University of Defense Technology,
Hunan 410073, China$^a$}
\address{and CCAST (World Laboratory) P.O. Box 8730, Beijing, 100080, China$^b$}
\date{January, 2002}
\maketitle
\thispagestyle{empty}
\vspace{15mm}
\begin{abstract}
We use the QCD sum rule approach to calculate the masses of the
$\Lambda_Q$ and $\Sigma_Q$ baryons to the $\Lambda_{QCD}/m_Q$
order within the framework of heavy quark effective theory. We
compare the direct approach and the covariant approach to this
problem. Two forms of currents have been adopted in our
calculation and their effects on the results are discussed.
Numerical results obtained in both direct and covariant approaches
are presented. The splitting between spin $1/2$ and $3/2$ doublets
derived from our calculation is ${\Sigma_Q^*}^2-\Sigma_Q^2\simeq
0.35\pm 0.03\,\mbox{GeV}^2$ which is in good agreement with the
experiment.
\end{abstract}
\vspace{4mm} 

\vspace{3.cm}
%\noindent
%December 2001
\newpage
\pagenumbering{arabic}
%%%%%%%%%%%%%%%%%%%%%%%%%%%%%%%%%%%%%%%%%%%%%%%%%%%%%%%%%%%%%%%%%%%%%%%%%%%%%%%%
\section{Introduction}
\label{sec1}

Important progresses in the theoretical description of hadrons containing a
heavy quark have been achieved with the development of the Heavy Quark
Effective Theory (HQET) \cite{HQET,review,KKP}. Based on the spin-flavor
symmetry of QCD, exactly valid in the infinite $m_Q$ limit, this framework
provides a systematic expansion of heavy hadron spectra and both the strong and
weak transition amplitudes in terms of the leading contribution, plus
corrections decreasing as powers of $1/m_Q$. HQET has been applied successfully
to learn about the properties of mesons and baryons made of both heavy and
light quarks.

The effective Lagrangian of the HQET, up to order $1/m_Q$, can be written as
\begin{eqnarray}
\label{Leff}
   {\cal L}_{\rm eff} = \bar h_v\,i v\!\cdot\!D\,h_v
   + \frac{1}{2 m_Q}\,{\cal K}
   + \frac{1}{2 m_Q}\,{\cal S}+{\cal O}(1/m_Q^2) \,,
\end{eqnarray}
where $h_v(x)$ is the heavy quark field in effective theory. Apart
from leading contribution, the Lagrangian density contains to
${\cal O}(1/m_Q)$ accuracy two additional operators $\cal K$ and
$\cal S$. ${\cal K}= \bar h_v\,(i D^\perp )^2\,h_v$ is the
non-relativistic kinetic energy operator and ${\cal S}=
\frac{1}{2}\,\left(\frac{\alpha_s(m_Q)}{\alpha_s(\mu)}
\right)^{3/\beta_0}\bar h_v\,\sigma_{\mu\nu} g_s\,G^{\mu\nu}\,h_v
$ is the chromo-magnetic interaction. Here $
(D^\perp)^2=D_\mu\,D^\mu-(v\cdot D)^2$, with
$D_\mu=\partial_\mu\,-i\,g A_\mu$ the covariant derivative and
$\beta_0=11-\frac{2}{3}n_f$ is the first coefficient of the
$\beta$ function.

The matrix elements of the operators $\cal K$ and $\cal S$ in
(\ref{Leff}) play a most significant role in many phenomenological
applications such as the spectroscopy of heavy hadrons \cite{FaNe}
and the description of inclusive decay rates \cite{Bigi}. For the
ground-state $\Lambda_Q$ and $\Sigma_Q$ baryons, one defines two
hadronic parameters, $\lambda_1$ and $\lambda_2$, as
\begin{eqnarray}
\label{matri}
\langle B(v)\mid {\cal K}\mid B(v)\rangle&=&\lambda_1,\nonumber\\
\langle B(v)\mid {\cal S}\mid B(v)\rangle&=&d_M\,\lambda_2.
\end{eqnarray}
where $d_M$ is zero for $\Lambda_Q$ and $-\frac{1}{2}\,$, $1\,$
for $\Sigma_{Q}^{*}$, $\Sigma_Q$ baryons, respectively. The
constant $d_M$ characterizes the spin-orbit interaction of the
heavy quark and the gluon field. Therefore, the mass of heavy
baryon up to order $1/m_Q$ corrections can be written in a compact
form
\begin{equation}
\label{mass} M=m_Q+\bar\Lambda-\frac{1}{2 m_Q}(\lambda_1+d_M\,\lambda_2)\;,
\end{equation}
where parameter $\bar\Lambda$ is the energy of the light degrees of freedom in
the infinite mass limit. Thus the splitting of the spin $1/2$ and $3/2$
doublets is
\begin{equation}
\Sigma_{Q}^{*\,2}-\Sigma_{Q}^{2}=\frac{3}{2}\;\lambda_2\;.
\end{equation}

The hadronic parameters $\lambda_1$ and $\lambda_2$ are nonperturbative ones
that should be either determined phenomenologically from experimental data or
estimated in some nonperturbative approaches. A viable approach is the QCD sum
rules \cite{svzsum} formulated in the framework of HQET \cite{hqetsum}. This
method allows us to relate hadronic observables to QCD parameters {\it via} the
operator product expansion (OPE) of the correlator. In the case of heavy
mesons, those two matrix elements thus masses had been calculated to perfect
first by Ball and Braun \cite{BaBr} and latterly by Neubert \cite{Neubert96}
taking a different approach. Masses of excited meson states had been calculated
up to $1/m_Q$ order in \cite{Dai97}. For the case of heavy baryons, there are
several attempts to calculate the baryonic matrix elements of ${\cal K}$ and
${\cal S}$. Using HQET sum rules, Colangelo $et.$ $al.$ have derived the value
of $\lambda_1$ for $\Lambda_Q$ baryon \cite{CDNP}. Furthermore, the baryonic
parameters $\lambda_1$ and $\lambda_2$ for the ground state baryons had been
calculated in \cite{Dai96} by evaluating the two-point correlation functions.
The mass parameters of the lowest lying excited heavy baryons had also been
determined recently in \cite{HuZZ}. In the present work we shall calculate the
baryonic parameters $\lambda_1$ and $\lambda_2$ for ground state $\Lambda_Q$
and $\Sigma_Q$ baryons using QCD sum rules in the HQET. Following Ball $et.$
$al.$ \cite{BaBr} and Neubert's \cite{Neubert96} work done for the meson, we
adopt these two approaches, named as direct approach and covariant approach, to
evaluate the three-point correlators and obtain the values of baryonic
parameters. It is of interest to compare the two methods in the analysis.

The remainder of this paper is organized as follows. In Sec. \ref{ssec1} we
introduce the interpolating currents for baryons and briefly present the
two-point sum rules. The direct Laplace sum rules analysis for the matrix
elements is presented in Sec. \ref{ssec2}. Another feasible approach (covariant
approach) to this aim can be found in Sec. \ref{ssec3}. Sec. \ref{sec3} is
devoted to numerical results and our conclusions. Some comments are also
available in Sec. \ref{sec3}.

\section{derivation of the sum rules for $\lambda_1$ and $\lambda_2$}
\label{sec2}
\subsection{Heavy baryonic currents and two-point sum rules}
\label{ssec1} The basic points in the application of QCD sum rules to problems
involving heavy baryon are to choose a suitable interpolating current in terms
of quark fields and to define the corresponding vacuum-to-baryon matrix
element. As is well known, the form of interpolating currents for baryon with
given spin and parity is not unique \cite{Ioff,GrYa,Ivanov}, the choice of
which one is just a question of predisposition. The most generally used form of
the heavy baryon current can be written as \cite{GrYa}
\begin{equation}
\label{cr} j^{\,v}=\epsilon_{abc}(q^{T\,a}_1\,C\Gamma \,\tau
\,q^{b}_2)\Gamma^\prime h_{v}^{c},
\end{equation}
in which $C$ is the charge conjugation matrix, $\tau$ is the
flavor matrix which is antisymmetric for $\Lambda_Q$ baryon and
symmetric for $\Sigma_{Q}^{(*)}$ baryon, $\Gamma$ and
$\Gamma^\prime$ are some gamma matrices, and a, b, c denote the
color indices. $\Gamma$ and $\Gamma^\prime$ can be chosen
co-variantly as
\begin{equation}
\label{lq} \Gamma=\gamma_5\;,  \hspace{1cm} \Gamma^\prime=1\;,
\end{equation}
for $\Lambda_Q$ baryon, and
\begin{eqnarray}
\label{sq} &&{}\Gamma=\gamma_\mu\;,\hspace{1cm} \Gamma^\prime=
(\gamma_\mu+v_\mu)\,\gamma_5\;,
\end{eqnarray}
for $\Sigma_Q$ baryon, and
\begin{eqnarray}
\label{sqs} \Gamma=\gamma_\nu\;,\hspace{0.5cm} \Gamma^\prime=-g_{\mu
\nu}+\frac{1}{3}\gamma_\mu\,\gamma_\nu -
\frac{1}{3}(\gamma_\mu\,v_\nu-\gamma_\nu\,v_\mu)+\frac{2}{3}v_\nu\,v_\mu\;,
\end{eqnarray}
for $\Sigma_{Q}^{*}$ baryon. Also the choice of $\Gamma$ is not unique. We can
insert a factor $\rlap/v$ before $\Gamma$ defined by equations
(\ref{lq})-(\ref{sqs}). The currents given by Eqs.(\ref{lq})-(\ref{sqs}) are
denoted as $j^{v}_{1}$ and that with $\rlap/v$ insertion as $j^{v}_{\,2}$,
which are two independent current representations.

The baryonic coupling constants in HQET are defined as follows
\begin{eqnarray}
\label{cc}
 \langle 0\mid j^v \mid
\Lambda(v)\rangle&=&F_{\Lambda}\,u,\nonumber\\
 \langle 0\mid j^v \mid
 \Sigma(v)\rangle&=&F_{\Sigma}u,\nonumber\\
 \langle 0\mid j^v \mid
\Sigma^*(v)\rangle&=&\frac{1}{\sqrt{3}}\,F_{\Sigma^*}u^{\alpha},
\end{eqnarray}
where $u$ is the spinor and $u_\alpha$ is the Rarita-Schwinger spinor in the
HQET, respectively. The coupling constants $F_{\Sigma}$ and $F_{\Sigma}^{*}$
are equivalent since $\Sigma_Q$ and $\Sigma_Q^*$ belong to the doublet with the
same spin-parity of the light degrees of freedom.

The QCD sum rule determination of these coupling constants can be done by
analyzing the two-point function
\begin{equation}
\label{twop} i \int dx e^{ik\cdot x}\langle 0\mid T\{j^v(x)\bar
j^v(0)\}\mid 0\rangle=\frac{1+\rlap/v}{2}Tr[\tau \tau^+]\Pi
(\omega),
\end{equation}
where $k$ is the residual momentum and $\omega=2v\cdot k$. It is
straightforward to obtain the two-point sum rule:
\begin{eqnarray}
F^{2}_{\Lambda}\;e^{-2\bar\Lambda_{\Lambda}/T}&=&\frac{3\,T^6}{2^5\,\pi^4}\,\delta_5(\omega_c/T)
+\frac{T^2}{2^7\,\pi^2}\,\langle\frac{\alpha_s}{\pi}G^2\rangle\,\delta_1(\omega_c/T)
+\frac{1}{6}\,\langle \bar q q\rangle^2,\nonumber\\
F^{2}_{\Sigma}\;e^{-2\bar\Lambda_{\Sigma}/T}&=&\frac{9\,T^6}{2^5\;\pi^4}\,\delta_5(\omega_c/T)
-\frac{T^2}{2^7\,\pi^2}\,\langle\frac{\alpha_s}{\pi}G^2\rangle\,\delta_1(\omega_c/T)+\frac{1}{2}\,\langle
\bar q q\rangle^2.
\end{eqnarray}
The functions $\delta_n(\omega_c/T)$ arise from the continuum
subtraction and are given by
\begin{equation}
   \delta_n(x) = \frac{1}{n!}\int\limits_0^x\!\mbox{d}t\,
   t^n e^{-t} = 1 - e^{-x} \sum_{k=0}^n \frac{x^k}{k!} \,.
\end{equation}
The second term of the last equation is assigned to the continuum
mode, which can be much larger than the ground state contributions
for the typical value of parameter $T$ due to the high dimensions
of the spectral densities.

%%%%%%%%%%%%%%%%%%%%%%%%%%%%%%%%%%%%%%%%%%%%%%%%%%%%%%%%%%%%%%%%%%%%%%
\subsection{The direct approach}\label{ssec2}
In order to evaluate the matrix elements $\lambda_1$ and $\lambda_2$ we
consider the three point correlation functions with $\cal{K}$ and $\cal{S}$
inserted directly between two interpolating currents at zero recoil as below
\begin{eqnarray}
i^2\int dx\int dy e^{i k\cdot x-i k'\cdot y}\langle 0\mid
T\{j^{\,v}(x){\cal K}(0)\bar j^{\,v}(y)\}\mid 0\rangle
=\frac{1+\rlap/v}{2}Tr[\tau \tau^+]\;T_K(\omega,\omega')\;,\nonumber\\
i^2\int dx\int dy e^{i k\cdot x-i k'\cdot y}\langle 0\mid T\{j^{\,v}(x){\cal
S}(0)\bar j^{\,v}(y)\}\mid 0\rangle =d_M\frac{1+\rlap/v}{2}Tr[\tau
\tau^+]\;T_S(\omega,\omega')\;,\label{ms}
\end{eqnarray}
where the coefficients $T_K(\omega,\omega')$ and $T_S(\omega,\omega')$ are
analytic functions in the ``off-shell energies" $\omega=2v\cdot k$ and
$\omega'=2v\cdot k'$ with discontinuities for positive values of these
variables. Saturating the three-point functions with complete set of baryon
states, one can isolate the part of interest, the contribution of the
lowest-lying baryon states associated with the heavy-light currents, as one
having poles in both the variables $\omega$ and $\omega'$ at the value
$\omega=\omega'=2\bar\Lambda$:
\begin{eqnarray}
T_K(\omega,\omega')=4\;\frac{\lambda_1F^2}{(2\bar
\Lambda-\omega)(2\bar\Lambda-\omega')}+\cdots,\nonumber\\
T_S(\omega,\omega')=4\;\frac{\lambda_2F^2}{(2\bar
\Lambda-\omega)(2\bar\Lambda-\omega')}+\cdots,\label{dpole}
\end{eqnarray}
where the ellipses denote the contribution of higher resonances. In the
theoretical calculation of the correlator it is convenient to choose the
residual momenta $k$ and $k^\prime$ parallel to the $v$, such that
$k_\mu=\frac{\omega}{2}v_\mu$ and $k'_\mu=\frac{\omega'}{2}v_\mu$.

The leading contribution to the matrix element of kinetic energy is of order
$1$, whereas to the chromo-magnetic interaction is of order $\alpha_s$.
Confining us to take into account these leading contributions of perturbation
and the operators with dimension $D\leq 6$ in OPE, the relevant diagrams in our
calculation are shown in Fig. 1. and Fig. 2. The calculation of the diagram (a)
in Fig. 2. is the most tedious one. It can be computed using Feynman
parameterization and the integral representation of the propagators, which is
the standard technique\cite{2loop,CoFP}. The factorization approximation has
been used to reduce the four-quark condensates to $\langle\bar qq\rangle^2$ in
the calculation.

On theoretical side the correlators $T_K(\omega,\omega')$ and
$T_S(\omega,\omega')$ can be casted into the form of integrals of
the double spectral densities as
\begin{eqnarray}
T_K(\omega, \omega')&=&\int\int
\frac{\mbox{d}s}{s-\omega}\frac{\mbox{d}s'}
{s'-\omega'}\rho_K(\mbox{s},\mbox{s}'),\nonumber\\
T_S(\omega,\omega')&=&\int\int
\frac{\mbox{d}s}{s-\omega}\frac{\mbox{d}s'}
{s'-\omega'}\rho_S(\mbox{s},\mbox{s}'),
\end{eqnarray}
where the double spectral density functions are
\begin{eqnarray}
&&\rho_{K}^{\Lambda,1}(s,s')=-\frac{3^3}{2^4\pi^4\,7!}s^7\delta(s-s')-
\frac{7}{2^6\pi^2\,3!}\langle\frac{\alpha_s}{\pi}G^2\rangle\,s^3\delta(s-s'), \nonumber\\
&&\rho_{K}^{\Sigma,1}(s,s')=-\frac{3^2\,11}{2^4\pi^4\,7!}s^7\delta(s-s')-
\frac{11}{2^6\pi^2\,3!}\langle\frac{\alpha_s}{\pi}G^2\rangle\,s^3\delta(s-s'), \nonumber\\
&&\rho_{K}^{\Lambda,2}(s,s')=-\frac{3^2\,5}{2^4\pi^4\,7!}s^7\delta(s-s')+
\frac{1}{2^6\pi^2\,3!}\langle\frac{\alpha_s}{\pi}G^2\rangle\,s^3\delta(s-s'), \nonumber\\
&&\rho_{K}^{\Sigma,2}(s,s')=-\frac{3^2}{\pi^4\,7!}s^7\delta(s-s')-
\frac{19}{2^6\pi^2\,3!}\langle\frac{\alpha_s}{\pi}G^2\rangle\,s^3\delta(s-s'), \nonumber\\
&&\rho_{S}^{\Sigma}(s,s\prime)=\frac{\alpha_s}{2^3\pi^5}[\Theta(s-s')
\int_{0}^{s'}\mbox{d}x\,(s-x)(s'-x)x^3+\Theta(s'-s)\int_{0}^{s}\mbox{d}x(s-x)(s'-x)x^3]
\nonumber\\ &&{}\qquad\qquad
-\frac{1}{48\pi^2}\langle\frac{\alpha_s}{\pi}G^2\rangle\,
s^3\delta(s-s')+\frac{4\alpha_s}{3\pi}\langle \bar q
q\rangle^2\;[s\delta(s')+s'\delta(s)].
\end{eqnarray}
The unitary normalization of flavor matrix $Tr[\tau\tau^+]=1$ has been applied
to get those densities. Here we use the numbers $1,2$ to denote the results
corresponding to the different choice of currents $j^{v}_{1}$ and $j^{v}_{2}$.
Those we do not discriminate with numeric superscripts meas that with or
without $\rlap/v$ insertion the results are identical. Following
Refs.~\cite{IWfun,DHHL,Neubert92D}, we then introduce new variables
$\omega_+=\frac 12(\omega+\omega')$ and $\omega_-=\omega-\omega'$, perform the
integral over $\omega_-$, and employ quark--hadron duality to equate the
remaining integral over $\omega_+$ up to a ``continuum threshold'' $\omega_c$
to the Borel transform of the double-pole contribution in (\ref{dpole}). Then
following the standard procedure we resort to the Borel transformation
$B^{\omega}_\tau$, $B^{\omega^{\prime}}_{\tau^{\prime}}$ to suppress the
contributions of the excited states. Considered the symmetries of the
correlation functions it is natural to set the parameters $\tau$,
$\tau^{\prime}$ to be the same and equal to $2T$, where $T$ is the Borel
parameter of the two-point functions. We end up with the set of sum rules
\begin{eqnarray}
-4\lambda_{1}^{\Lambda,\,1}F^2e^{-2\bar\Lambda_\Lambda/T}&=&
\frac{3^3\;T^8}{(2\;\pi)^4}\delta_7(\omega_c/T)
+\frac{7\;T^4}{2^6\;\pi^2}\langle\frac{\alpha_s}{\pi}G^2\rangle\,\delta_3(\omega_c/T),\nonumber\\
-4\lambda_{1}^{\Lambda,\,\,2}F^2e^{-2\bar\Lambda_\Lambda/T}&=&\frac{3^2\;
5\;T^8}{(2\;\pi)^4}\delta_7(\omega_c/T)
-\frac{T^4}{2^6\;\pi^2}\langle\frac{\alpha_s}{\pi}G^2\rangle\,\delta_3(\omega_c/T),\nonumber\\
-4\lambda_{1}^{\Sigma,\,1}F^2e^{-2\bar\Lambda_\Sigma/T}&=&\frac{3^2\;11\;T^8}{(2\;\pi)^4}
\delta_7(\omega_c/T)
+\frac{11\;T^4}{2^6\;\pi^2}\langle\frac{\alpha_s}{\pi}G^2\rangle\,\delta_3(\omega_c/T),\nonumber\\
-4\lambda_{1}^{\Sigma,\,2}F^2e^{-2\bar\Lambda_\Sigma/T}&=&\frac{3^2\;T^8}{\pi^4}\delta_7(\omega_c/T)
+\frac{19\;T^4}{2^6\;\pi^2}\langle\frac{\alpha_s}{\pi}G^2\rangle\,\delta_3(\omega_c/T),\nonumber\\
\;4\lambda_2F^2e^{-2\bar\Lambda_\Sigma/T}&=&\frac{12}{\pi^4}\frac{\alpha_s}{\pi}T^8\delta_7(\omega_c/T)
-\frac{T^4}{8\pi^2}\langle\frac{\alpha_s}{\pi}G^2\rangle\,\delta_3(\omega_c/T)+\frac{32\;T^2\alpha_s}{3\pi}\langle
\bar q q\rangle^2\;\delta_1(\omega_c/T),
\end{eqnarray}

It is worth noting that the next-to-leading order $\alpha_s$ corrections have
not been included in the sum rule calculations. However, the baryonic parameter
obtained from the QCD sum rules actually is a ratio of the three-point
correlator to the two-point correlator results.  While both of these
correlators are subject to large perturbative QCD corrections, it is expected
that their ratio is not much affected by these corrections  because of
cancellation. On the other hand, we have only calculated the diagonal sum rules
by using the same type interpolating current in the correlator. As to the
non-diagonal sum rules, the only non-vanishing contributions in the OPE of
correlator are terms with odd number of dimensions, thus the perturbative term
gives no contribution. The resulting sum rules are dominated by the quark-gluon
condensates. It is expected that the non-diagonal sum rules will give no more
information than diagonal ones. This has been proved to be true in the analysis
of Ref. \cite{GrKo}.

%%%%%%%%%%%%%%%%%%%%%%%%%%%%%%%%%%%%%%%%%%%%%%%%%%%%%%%%%%%%%%
\subsection{The covariant approach} \label{ssec3}
In the previous subsection we have completed the task of the
determination of the matrix elements for both the operators of
kinetic energy and chromo-magnetic interaction by direct
calculation of three-point correlation functions. In fact, there
exists a field-theory analog of the virial
theorem\cite{virial,BiSU} in consideration of the restrictions the
equation of motion and the heavy quark symmetry imposing on
baryons, which relates the kinetic energy and chromo-interaction
to each other and ensure the intrinsic smallness of the kinetic
energy explicitly. In this subsection we shall follow Neubert's
procedure\cite{Neubert96} and take those restrictions into account
to deduce a new result of the kinetic energy (the chromo-magnetic
interaction is identical).

The main idea of that procedure is that the coefficients of the covariant
decomposition of the bilinear matrix elements, the so called invariant
functions, can be related to the kinetic energy and chromo-magnetic interaction
at the zero recoil. Following the discussion in \cite{FaNe,Mannel} we have the
general decomposition (see Appendix):
\begin{equation}
\label{svl} \langle \Lambda\mid\bar h_v
\sigma_{\mu\nu}ig_sG^{\mu\nu}
h_{v^{\prime}}\mid\Lambda^{\prime}\rangle=
\phi_1(v',v)(v_{\mu}^{\prime}v_{\nu}-v_{\mu}v_{\nu}^{\prime}) \bar
u \sigma^{\mu\nu}u^{\prime},
\end{equation}
for $\Lambda_Q$ baryon, in which $u$ is the spinor in HQET, and
\begin{equation}
\label{svs} \langle \Sigma\mid\bar h_v
\sigma_{\mu\nu}ig_sG^{\mu\nu}
h_{v^{\prime}}\mid\Sigma^{\prime}\rangle=\phi^{\mu\nu}_{\alpha\beta}(v',v)
\bar\Psi^{\alpha}\sigma^{\mu\nu}
\Psi^{\prime\beta} ,
\end{equation}
for $\Sigma_Q$ baryon, where $\phi^{\mu\nu}_{\alpha\beta}$ bears
the decomposition
\begin{eqnarray}
\phi^{\mu\nu}_{\alpha\beta}&=&\phi_{1}(g_{\mu\alpha}g_{\nu\beta}-g_{\mu\beta}g_{\nu\alpha})
+\phi_{2}(g_{\mu\beta}v_{\nu}v^{\prime}_{\alpha}-g_{\nu\beta}v_{\mu}v^{\prime}_{\alpha}
+g_{\nu\alpha}v_{\beta}v^{\prime}_{\mu}-g_{\mu\alpha}v_{\beta}v^{\prime}_{\nu})\nonumber\\
&&{}+\phi_{3}(g_{\alpha\mu}v_{\nu}v_{\beta}-g_{\nu\alpha}v_{\mu}v_{\beta}
+g_{\nu\beta}v^{\prime}_{\alpha}v^{\prime}_{\mu}-g_{\mu\beta}v^{\prime}_{\alpha}v^{\prime}_{\nu})
\nonumber\\&&{}+\phi_{4}(v^{\prime}_{\mu}v_{\nu}-v^{\prime}_{\nu}v_{\mu})g_{\alpha\beta}
+\phi_{5}(v^{\prime}_{\mu}v_{\nu}-v^{\prime}_{\nu}v_{\mu})v^{\prime}_{\alpha}v_{\beta},\label{phidec}
\end{eqnarray}
in which we use the covariant representation of the doublets
$\Psi_{\mu}=u_{\mu}+\frac{1}{\sqrt{3}}(v_{\mu}+\gamma_{\mu})u$
with restrictions $\rlap/v u=u$, $v_{\mu}u_{\mu}=0$ and
$\gamma_{\mu}u_{\mu}=0$. The normalization of those coefficients
at zero recoil is $ \phi_1(1)=-\frac{1}{3}\lambda_1$ for
$\Lambda_Q$ baryon and
\begin{eqnarray}
&&\pm\lambda_2=2\,\phi_1(1),\nonumber\\
\pm\lambda_1=\phi_{0}(1)&=&\phi_{1}(1)-2(\phi_{2}(1)-\phi_{3}(1))-3\phi_{4}(1),
\end{eqnarray}
for $\Sigma_Q$ baryon. The foregoing minus sign corresponds to the
$\Sigma_Q$ baryon and plus to $\Sigma_Q^{*}$ baryon.

Let us now derive the Laplace sum rules for the invariant functions
$\phi_i(w)$. The analysis proceeds in complete analogy to that of the
Isgur--Wise function. We shall only briefly sketch the general procedure and
refer for details to Refs.~\cite{IWfun,DHHL}. We consider, in the HQET, the
three-point correlation function of the local operator appearing in (\ref{svs})
with two interpolating currents for the ground-state heavy baryons:
\begin{eqnarray}
   &&i^2\;\int\mbox{d}x\,\mbox{d}y\,e^{ik\cdot x-ik'\cdot y}\,
    \langle\,0\,|\,\mbox{T}\left\{j^{v}(x)\,,
    \bar h_{v} i\,g_s\Gamma\, G^{\mu\nu} h_{v'}(0) ,
    \bar j^{v'}\,(y) \right\} |\,0\,\rangle \nonumber\\
   &&\quad =\Phi^{\mu\nu}_{\alpha\beta}(v',v,k',k)
    \Gamma^{\prime}_{\alpha}\frac{1+\rlap/v}{2}\Gamma\;\frac{1+\rlap/v^{\prime}}{2}
\bar\Gamma^{\prime}_{\beta}\,, \label{correl}
\end{eqnarray}
where $k$ and $k'$ are the residual momenta. % In the theoretical
%calculation we take them parallel to the $v$ and $v'$,respectively.
The Dirac structure of the correlation function, as shown in the
second line,  is a consequence of the Feynman rules of the HQET.
$\Phi^{\mu\nu}_{\alpha\beta}$ obeys a decomposition analogous to
(\ref{phidec}), with coefficient functions
$\Phi_i(\omega,\omega',w)$ that are analytic in the ``residual
energy'' $\omega=2 v\cdot k$ and $\omega'=2 v'\cdot k'$, with
discontinuities for positive values of these variables. These
functions also depend on the velocity transfer $w=v\cdot v'$.

The lowest-lying states are the ground-state baryons $B(v)$ and
$B'(v')$ associated with the heavy-light currents. They lead to a
double pole located at $\omega=\omega'= 2\bar\Lambda$. The residue
of this double pole is proportional to the invariant functions
$\phi_i(w)$. We find
\begin{equation}
   \Phi_i^{\rm pole}(\omega,\omega',w)
   = \frac{4\,s_c\,\phi_i(w)\,F^2}
   {(\omega-2\bar\Lambda)(\omega'-2\bar\Lambda)} \,,
\label{pole}
\end{equation}
where $s_c$ is the structure constant, $1$ for $\Lambda_Q$,
$-\;\frac{2+w}{9}$ for $\Sigma_Q$ and $1$ for $\Sigma_{Q}^{*}$
baryons. In the deep Euclidean region the correlation function can
be calculated perturbatively because of asymptotic freedom.
Following the standard procedure, we write the theoretical
expressions for $\Phi_i$ as double dispersion integrals and
perform a Borel transformation in the variables $\omega$ and
$\omega'$, then set the associated Borel parameters equal:
$\tau=\tau'\equiv 2 T$. All goes like that in the direct approach,
we introduce new variables $\omega_+=\frac 12(\omega+\omega')$ and
$\omega_-=\omega-\omega'$, perform the integral over $\omega_-$,
and get the Laplace sum rules at zero recoil:
\begin{eqnarray}
8s_cF^2\phi_1e^{-2\bar\Lambda_\Sigma/T}=\frac{4}{\pi^4}\frac{\alpha_s}{\pi}T^8\delta_7(\omega_c/T)
&-&\frac{1}{24\pi^2}\langle\frac{\alpha_s}{\pi}G^2\rangle\,\delta_3(\omega_c/T)+\frac{32\alpha_s}{9\pi}\langle
\bar q q\rangle^2\;T^2\delta_1(\omega_c/T),
\nonumber\\
8s_cF^2\phi_1e^{-2\bar\Lambda_\Sigma/T}&=&8s_cF^2\,2(\phi_2-\phi_3)e^{-2\bar\Lambda_\Sigma/T},
\nonumber\\
8s_cF^2\phi_4e^{-2\bar\Lambda_\Sigma/T}&=&\frac{2}{\pi^4}\frac{\alpha_s}{\pi}
T^8\delta_7(\omega_c/T),\nonumber\\
8s_cF^2\phi_5e^{-2\bar\Lambda_\Sigma/T}&=&-\frac{1}{\pi^4}\frac{\alpha_s}{\pi}
T^8\delta_7(\omega_c/T)
\end{eqnarray}
for $\Sigma_Q$ baryon, and
\begin{equation}
8F^2\phi_1e^{-2\bar\Lambda_\Lambda/T}=-\frac{2}{\pi^4}\frac{\alpha_s}{\pi}
T^8\delta_7(\omega_c/T),
\end{equation}
for $\Lambda_Q$ baryon. After some simple algebra we find
\begin{eqnarray}
&&\qquad\quad\quad-4\lambda_1F^2e^{-2\bar\Lambda_\Sigma/T}=\frac{9}{\pi^4}\frac{\alpha_s}{\pi}
T^8\delta_7(\omega_c/T),\nonumber\\
4\lambda_2F^2e^{-2\bar\Lambda_\Sigma/T}&=&\frac{12}{\pi^4}\frac{\alpha_s}{\pi}T^8\delta_7(\omega_c/T)
-\frac{T^4}{8\pi^2}\langle\frac{\alpha_s}{\pi}G^2\rangle\,\delta_3(\omega_c/T)+\frac{32\;T^2\alpha_s}{3\pi}\langle
\bar q q\rangle^2\;\delta_1(\omega_c/T),
\end{eqnarray}
for $\Sigma_Q$ baryon, and
\begin{equation}
-4\lambda_1F^2e^{-2\bar\Lambda_\Lambda/T}=-\frac{3T^8}{\pi^4}\frac{\alpha_s}{\pi}
\delta_7(\omega_c/T),
\end{equation}
for $\Lambda_Q$ baryon. The minus sign of the $\Lambda_Q$ baryon
result may seem bizarre, in Sec.\ref{sec3} we will return to dwell
on this point.
%%%%%%%%%%%%%%%%%%%%%%%%%%%%%%%%%%%%%%%%%%%%%%%%%%%%%%%%%%%%%%%%%%%%%%%%%%%
\section{numerical results and  conclusions}
\label{sec3} In order to get the numerical results, we divide our three-point
sum rules by two-point functions to obtain $\lambda_1$ and $\lambda_2$ as
functions of the continuum threshold $\omega_c$ and the Borel parameter $T$.
This procedure can eliminate the systematic uncertainties and cancel the
parameter $\bar\Lambda$. As for the condensates, we adopt the standard values
\begin{eqnarray}
&&\langle\bar q q\rangle = -(0.23\;\mbox{GeV})^3,\nonumber\\
&&\langle\frac{\alpha_s}{\pi}G^2\rangle = 0.012\;\mbox{GeV}^4.
\end{eqnarray}
From two-point sum rules one has known that there exist stable
windows between $0.8<T<1.2\;\mbox{GeV}$ and
$\omega_c=(2.2-2.7)\;\mbox{GeV}$ for $\Lambda_Q$ baryon, and
$0.7<T<1.1\;\mbox{GeV}$ and $\omega_c=(2.6-3.3)\;\mbox{GeV}$ for
$\Sigma_Q$ baryon. The stability window for three-point functions
starts almost from values of the Borel parameter at
$0.7\;\mbox{GeV}$ and stretches practically to
$T\rightarrow\infty$. It is known that stability at lager Borel
parameter could not give any valuable information, since in this
region the sum rule is strongly effected by the continuum model.
The usual criterium that both the higher-order power corrections
and the continuum contribution should not be very large restricts
the working region considerably. In the case of the three-point
functions, the results are severely smeared by the continuum
contributions for the high dimension of spectral densities and
thus it is very difficult to ensure the contribution of the
continuum mode is small. The working region of the three-point
functions should be determined by the stable region of the
two-point functions. So it does not necessarily coincide with the
stable windows of the three-point functions \cite{BaBr}. Thus we
find our working region for three-point functions is
$T=0.8-1.2\;\mbox{GeV}$ for $\Lambda_Q$ baryon and
$T=0.7-1.1\;\mbox{GeV}$ for $\Sigma_Q$ baryon. The results for
$\lambda_1$ of the direct approach with two different choices of
currents are shown in Fig. 3 for $\Lambda_Q$ baryon and Fig. 4 for
$\Sigma_Q$ baryon. Results for kinetic energy of $\Lambda_Q$ and
$\Sigma_Q$ baryons obtained by the covariant approach are
presented in Fig. 5. The forms of the chromo-magnetic interaction
obtained by both approaches do not differ from each other, so we
plot that unique curve in one figure, Fig. 6. For the $\Lambda_Q$
baryon we obtain the residual mass $\bar\Lambda_\Lambda=0.8\pm
0.1\;\mbox{GeV}$ and
\begin{eqnarray}
-\lambda_{1}^{1}=0.4\pm 0.1\;\mbox{GeV}^2,\nonumber\\
-\lambda_{1}^{2}=0.5\pm 0.1\;\mbox{GeV}^2,
\end{eqnarray} in the direct approach, where the superscripts denote the
different choices of currents,and \begin{equation}
-\lambda_1=-(0.08\pm 0.02)\;\mbox{GeV}^2,
\end{equation} in the covariant approach.
  For the $\Sigma_Q$ baryon the effective mass we obtained is
$\bar\Lambda_\Sigma=1.0\pm 0.1\;\mbox{GeV}$ and
\begin{eqnarray}
-\lambda_{1}^{1}=0.7\pm 0.2\;\mbox{GeV}^2,\nonumber\\
-\lambda_{1}^{2}=1.0\pm 0.2\;\mbox{GeV}^2,
\end{eqnarray} in the direct approach, where the superscripts also
denote different currents, and
\begin{equation}
-\lambda_1=0.11\pm 0.03\;\mbox{GeV}^2,
\end{equation} in covariant approach. For the chromo-magnetic
interaction for $\Sigma$ baryon the results read \begin{equation}
\lambda_2=0.23\pm 0.02\;\mbox{GeV}^2.
\end{equation}Then we get the splitting of the spin $1/2$ and spin
$3/2$ doublets is
\begin{equation}
\Sigma^{*2}_Q-\Sigma^2_Q=\frac{3}{2}\;\lambda_2=0.35\pm
0.03\;\;\mbox{GeV}^2.
\end{equation}All error quoted before is due to the variation of
Borel parameter $T$ and the continuum threshold $\omega_c$. When
it is scaled up to the bottom quark mass scale there will be a
factor $\sim 0.8$ approximately due to the renormalization group
improvement.

As for the effects on the correlation function of the different
choices of the interpolating currents we may assert some facts and
inspections. From the preceding numerical results it is clear that
the interpolating currents with the $\rlap/ v$ insertion give a
considerable larger result to the kinetic energy than those
without the insertion. Nevertheless, the two-point sum rules do
not differ with the different currents\cite{Dai96,GrYa}. From our
calculation it is explicit that the sum rules associated with
chromo-interaction insertion are identical. In our covariant
calculation we find that the invariant functions do differ from
each other generally, but interestingly they coincide at the zero
recoil thus the chromo-interaction and kinetic energy do not take
a different form. Naively, we can tell that the disparity of the
two forms of the kinetic energy obtained in our direct calculation
mainly comes from the Lorenz structural differences of the two
interpolating currents. It may be noted that derivative operator
acts differently on the currents with or without $\rlap/ v$
insertion, thus with the insertion it is easier for the continuum
contamination to go into the correlation functions. It is urgently
needed to exclude the continuum contribution which smeared heavily
the results of both the direct and covariant approaches. It is
this continuum contamination that makes the prediction of the
kinetic energy more intriguing. All previous theoretical
calculations with QCD sum rule approach or lattice calculation
give various results and can differ from each other by several
times \cite{CDNP,Dai96,BaBr,Neubert96,Neubert92D,GiMS}. Current
experimental data is not enough to judge which one is right and
what we can get is some restrictions on the kinetic
energy\cite{Mannel94} or a rough estimate of the kinetic energy
extracted from experimental data with some assumption\cite{JeKi}.
As demonstrated in Refs.~\cite{BiSU,BSUV95} using a toy model of
harmonic oscillator, the main origin of the discrepancy between
the direct and covariant approach is the continuum smeared
contribution. In the direct approach the first excited
contribution plays an important role. If we want to suppress this
contribution, we go to such a large Borel parameter that the power
corrections blow up. For acceptable Borel parameter, we get an
over-estimated sum rule for the the kinetic energy. In the
covariant approach (via virial theorem) the situation is
especially bad. The excited contribution consists of two
components -- the diagonal transitions and off-diagonal ones --
and each one is large, but they have opposite relative sings. For
highly excited states the sign-alternating terms are smeared to
zero after summation. However, the first two terms do not cancel
with each other and screen the ground state contribution. Thus a
lower-estimated result will be obtained. The minus sign before the
kinetic energy of $\Lambda_Q$ baryon in the covariant approach may
be seen as a manifestation of this assertion. Due to the unknown
weight, we cannot annihilate those contributions by weighted
averaging just like that in the Quantum Mechanics. But we may
safely take the results of the direct and covariant approach as
lower-bound and higher-bound of the kinetic energy parameter,
respectively. Then, follow\cite{Braun99}, to take the mean value
of the direct and covariant approach results as an rough estimate.
The result thus obtained is
\begin{eqnarray}\label{lambda1}
-\bar\lambda_1^1&\simeq &0.18\pm 0.06\;\mbox{GeV}^2 ,\nonumber\\
-\bar\lambda_1^2&\simeq &0.24\pm 0.06\;\mbox{GeV}^2,
\end{eqnarray}
for $\Lambda_Q$ baryon and
\begin{eqnarray}
-\bar\lambda_1^1&\simeq & 0.39\pm 0.12\;\mbox{GeV}^2,\nonumber\\
-\bar\lambda_1^2&\simeq & 0.54\pm 0.12\;\mbox{GeV}^2,
\end{eqnarray}
for $\Sigma_Q$ baryon. Taking all results obtained the mass of the ground state
baryon is on hand. From $m_{\Lambda_c}$ and $m_{\Lambda_b}$\cite{data}, we
determine the heavy quark masses $m_c\simeq 1.41\pm 0.16\;\mbox{GeV}$ and
$m_b\simeq 4.77\pm 0.12\;\mbox{GeV}$. In the determination we have taken the
average of the results obtained from two interpolating currents to be the
physical pole masses of the heavy quarks because that the difference of the
corresponding mass does not exceed the error bar. These values give the
following results:
\begin{eqnarray}
m_{\Sigma_c}\simeq 2.47\pm 0.20\;\mbox{GeV},\nonumber\\
m_{\Sigma_c^*}\simeq 2.59\pm 0.20\;\mbox{GeV},\nonumber\\
m_{\Sigma_b}\simeq 5.79\pm 0.13\;\mbox{GeV},\nonumber\\
m_{\Sigma_b^*}\simeq 5.82\pm 0.13\;\mbox{GeV},
\end{eqnarray} with interpolating current $j_{v}^{1}$ and
\begin{eqnarray}
m_{\Sigma_c}\simeq 2.52\pm 0.20\;\mbox{GeV},\nonumber\\
m_{\Sigma_c^*}\simeq 2.64\pm 0.20\;\mbox{GeV},\nonumber\\
m_{\Sigma_b}\simeq 5.80\pm 0.13\;\mbox{GeV},\nonumber\\
m_{\Sigma_b^*}\simeq 5.83\pm 0.13\;\mbox{GeV},
\end{eqnarray} with interpolating current $j_{v}^{2}$.
The spin average of the doublets are free of the chromo-interaction
contribution and thus free of the uncertainties involved in the calculation of
$\lambda_2$. Average over the doublets we have the quantity
\begin{eqnarray}
\frac{1}{3}\;(M_{\Sigma_Q}
+2M_{\Sigma_{Q}^{*}})=m_Q+\bar\Lambda_\Sigma +\frac{1}{2
m_Q}\bar\lambda_1 \nonumber \end{eqnarray} which is more reliable.
For the $c$ quark case, it is $2.55\pm 0.20\;\mbox{GeV}$ with
current $j_{v}^{1}$ and $2.60\pm 0.20\;\mbox{GeV}$ with current
$j_{v}^{2}$. For the $b$ quark case it is $5.81\pm
0.13\;\mbox{GeV}$ with current $j_{v}^{1}$ and $5.83\pm
0.13\;\mbox{GeV}$ with current $j_{v}^{2}$. Experimentally
$M_{\Sigma_c}=2453\pm 0.2\;\mbox{MeV}$\cite{data}. There is
experimental evidence for $\Sigma_{c}^{*}$ at
$M_{\Sigma_{c}^{*}}=2519\pm 2\;\mbox{MeV}$\cite{cleo}. If we take
this value for $\Sigma_{c}^{*}$, we have
$\frac{1}{3}(M_{\Sigma_c}+2\;M_{\Sigma_{c}^{*}})=2497\pm
1.4\;\mbox{MeV}$ which is in reasonable agreement with the
theoretical prediction. For lack of experimental data the
corresponding quantity for the bottom quark will be checked in the
future. If we take the preceding masses of the charmed $\Sigma$
baryons the splitting thus reduced is $0.33\;\mbox{GeV}^2$ and our
theoretical splitting is in considerable agreement with the
experimental data.

As the kinetic energy of $\Lambda_Q$ baryon can be related to the
spectrum via the kinetic energy of meson
as\cite{spectrum}\footnote{The relation between $\mu^{2}_{\pi}$ in
Ref.~\cite{spectrum} and $\lambda_1$ in this paper is
$\mu^{2}_{\pi}=-\lambda_1$}
\begin{equation}
(m_{\Lambda_b}-m_{\Lambda_c})-(\overline{m}_B-\overline{
m}_D)=[\lambda_1(\Lambda_b)-\lambda_1(B)]\left(\frac{1}{2m_c}-\frac{1}{2m_b}\right)
+{\cal{O}}(1/m^{2}_{Q})\;,
\end{equation}
where $\overline{ m}_B=\frac{1}{4}(m_B+3m^{*}_{B})$ and $\overline{
m}_D=\frac{1}{4}(m_D+3m^{*}_{D})$ denote the spin-averaged meson masses, the
difference between the kinetic energy of $B$ meson and that of $\Lambda_b$
baryon can be extracted as
\begin{eqnarray}
\lambda_1(\Lambda_b)-\lambda_1(B)= 0.01\pm 0.02\;\mbox{GeV}^2\;,
\end{eqnarray}
which is consistent with the value obtained in Ref.~\cite{spectrum}. Resorting
to the recent experimental data for the mesonic kinetic energy parameter
obtained in the inclusive semileptonic $B$ decays\cite{cleo-exp},
$-\lambda_1=0.24\pm 0.11\;\mbox{GeV}^2$, one can thus get the value of baryonic
kinetic energy as
\begin{eqnarray}
-\lambda_1(\Lambda_b)=0.23\pm 0.13\;\mbox{GeV}^2\;,
\end{eqnarray}
which is in reasonable agreement with our theoretical prediction given in
(\ref{lambda1}).

For conclusions, we have calculated the $1/m_Q$ corrections to the heavy baryon
masses from the QCD sum rules within the framework of the HQET. Two approaches
have been adopted in the evaluation of the three-point correlators. Our final
results read
\begin{eqnarray}
M_{\Sigma_Q}&=&m_Q + \bar\Lambda_\Sigma + \frac{1}{2m_Q}(0.16\pm
0.12\;\mbox{GeV}^2),\nonumber\\
M_{\Sigma_{Q}^{*}}&=&m_Q + \bar\Lambda_\Sigma +
\frac{1}{2m_Q}(0.51\pm 0.12\;\mbox{GeV}^2),
\end{eqnarray} for interpolating current without $\rlap/v$ insertion and
\begin{eqnarray}
M_{\Sigma_Q}&=&m_Q + \bar\Lambda_\Sigma + \frac{1}{2m_Q}(0.31\pm
0.12\;\mbox{GeV}^2),\nonumber\\
M_{\Sigma_{Q}^{*}}&=&m_Q + \bar\Lambda_\Sigma +
\frac{1}{2m_Q}(0.66\pm 0.12\;\mbox{GeV}^2),
\end{eqnarray}
for interpolating current with $\rlap/v$ insertion. The $1/m_Q$ corrections are
small. We have taken the mean value of the direct and covariant approach as the
rough estimate of the kinetic energy parameter $\lambda_1$. Our theoretical
predictions are in agreement with the recent experimental data. For a more
precise treatment of the kinetic energy, more sophisticated technique to
distinguish the smearing continuum contribution is in urgent necessity to be
developed.
%%%%%%%%%%%%%%%%%%%%%%%%%%%%%%%%%%%%%%%%%%%%%%%%%%%%%%%%%
\acknowledgments This work is supported in part by the National Natural Science
Foundation of China under Contract No. 19975068.
\vspace{0.5cm}
%%%%%%%%%%%%%%%%%%%%%%%%%%%%%%%%%%%%%%%%%%%%%%%%%%%%%%%%
\begin{appendix}
\section{the decomposition of the bilinear matrix element}
In this appendix, we present the decomposition of bilinear matrix
element. There exists the decomposition of $\Lambda_Q$ baryon, we
present it here merely for completeness and convention.

First, let us consider the bilinear matrix element over
$\Lambda_Q$ baryons
\begin{equation}
\label{matla} \langle \Lambda\mid\bar h_v(-i
\overleftarrow{D_{\mu}})\Gamma^{\mu\nu}
iD_{\nu}h_{v^{\prime}}\mid\Lambda^{\prime}\rangle=\psi_{\mu\nu}(v^{\prime},v)\bar
u\Gamma^{\mu\nu}u^{\prime},
\end{equation}
the coefficients obey the symmetric relation
$\psi_{\mu\nu}(v^{\prime},v)=\psi_{\nu\mu}^{*}(v,v^{\prime})$. It
is convenient to write the coefficient $\psi$ into the sum of
symmetric and anti-symmetric parts
$\psi_{\mu\nu}=\frac{1}{2}\;[\psi_{\mu\nu}^{A}+\psi_{\mu\nu}^{S}]$
which can be presented covariantly as
\begin{eqnarray}
\label{psila}
\psi_{\mu\nu}^{A}&=&\psi_{1}^{A}(v_{\mu}^{\prime}v_{\nu}-v_{\mu}v_{\nu}^{\prime}),\nonumber\\
\psi_{\mu\nu}^{S}&=&\psi_{1}^{S}g_{\mu\nu}+\psi_{2}^{S}(v+v^{\prime})_{\mu}(v+v^{\prime})_{\nu}
+\psi_{3}^{S}(v-v^{\prime})_{\mu}(v-v^{\prime})_{\nu},
\end{eqnarray}
the HQET equation of motion implies that
$v^{\prime}_{\nu}\psi_{\mu\nu}=0$ from which we can obtain the
relations between those coefficients
\begin{eqnarray}
\psi_{1}^{S}+(1+y)\psi_{2}^{S}+(1-y)\psi_{3}^{S}+y\psi_{1}^{A}&=&0,\nonumber\\
(1+y)\psi_{2}^{S}+(y-1)\psi_{3}^{S}-\psi_{1}^{A}&=&0,\label{psirela}
\end{eqnarray}
with\begin{equation} \label{deria} \bar h\,i\,\overleftarrow{
D_{\mu}}\Gamma h^{\prime}+\bar h\,i\,D_{\mu}\Gamma h^{\prime}
=i\,\partial_{\mu}(\bar h\,\Gamma\,h^{\prime})
\end{equation}
bear in mind we can get
\begin{equation}
\label{rmatla} \langle \Lambda\mid\bar h_v \Gamma^{\mu\nu}iD_{\mu}
iD_{\nu}h_{v^{\prime}}\mid\Lambda^{\prime}\rangle=\psi_{\mu\nu}(v^{\prime},v)\bar
u\Gamma^{\mu\nu}u^{\prime}+\bar\Lambda
(v^{\prime}-v)_{\mu}\xi_{\nu}\bar u\Gamma^{\mu\nu}u^{\prime},
\end{equation}
using $x$ dependence of state in HQET $\mid B(x)\rangle=e^{-i
\bar\Lambda v\cdot x}\mid B(0)\rangle$. The $\xi_{\nu}$ are
defined as
\begin{equation}
\label{xila} \langle \Lambda\mid\bar h_v \Gamma^{\nu}
iD_{\nu}h_{v^{\prime}}\mid\Lambda^{\prime}\rangle=\xi_{\nu}\bar
u\Gamma^{\nu}u^{\prime},
\end{equation}
Similarly, we can define the matrix elements for the operators of
kinetic energy and chromo-interaction over baryon states with
different velocities
\begin{equation}
\label{svla} \langle \Lambda\mid\bar h_v
\sigma_{\mu\nu}iG^{\mu\nu}
h_{v^{\prime}}\mid\Lambda^{\prime}\rangle=\phi_1(v_{\mu}^{\prime}v_{\nu}-v_{\mu}v_{\nu}^{\prime})
\bar u \sigma^{\mu\nu}u^{\prime},
\end{equation}
\begin{equation}
\label{kvla} \langle \Lambda\mid\bar h_v (iD^{\perp})^2\Gamma
h_{v^{\prime}}\mid\Lambda^{\prime}\rangle=\phi_0 \bar u\Gamma
u^{\prime},
\end{equation}
once such defined, the $\psi_i$ can be expressed via two $\phi_i$
\begin{eqnarray}
\label{psi2phila}
\psi_{1}^{A}&=&\phi_1-\bar\Lambda^{2}\xi\frac{y-1}{y+1},\nonumber\\
\psi_{1}^{s}&=&\phi_0+y\phi_1+\bar\Lambda^{2}\xi\frac{y-1}{y+1},\nonumber\\
\psi_{3}^{s}&=&\frac{(1+2y)\phi_1+\phi_0}{2(y-1)}-\frac{y}{2(y+1)}\bar\Lambda^{2}\xi,\nonumber\\
\psi_{2}^{s}&=&\frac{\psi_{1}^{A}-(y-1)\psi_{3}^{S}}{1+y},
\end{eqnarray}
 the normalization of $\phi_0, \phi_1$ are
$\phi_0(1)=\lambda_1, \phi_1(1)=-\frac{1}{3}\phi_0(1)$, thus we
get the desired result.

Generalization can be made to the higher spin states such as
$\Sigma_{Q}$ baryon. The procedure goes almost the same. The only
difference lies on the decomposition of the matrix element. Hence
we will give the forms of decomposition and the final desired
relation, others we will not dwell on. The covariant
representation of the doublet is
$\Psi_{\mu}=u_{\mu}+\frac{1}{\sqrt{3}}(v_{\mu}+\gamma_{\mu})u$.
The matrix element is
\begin{equation}
\langle \Sigma\mid\bar
h_v(-i\overleftarrow{D_{\mu}})\Gamma^{\mu\nu}
iD_{\nu}h_{v^{\prime}}\mid\Sigma^{\prime}\rangle=\psi_{\mu\nu}^{\alpha\beta}(v^{\prime},v)\bar
\Psi_{\alpha}\Gamma^{\mu\nu}\Psi^{\prime}_{\beta},\label{matsa}
\end{equation}
in which the coefficients obey symmetric relation
$\psi_{\mu\nu}^{\alpha\beta}(v,v^{\prime})=\psi_{\nu\mu}^{\beta\alpha}(v^{\prime},v)
$. Adopt the same symmetric and antisymmetric decomposition of the
coefficients like that in $\Lambda_Q$ baryon case, we have
\begin{eqnarray}
\label{psisa} \psi_{\mu\nu}^{\alpha\beta,\,A}&=&
\psi_{1}^{A}(g_{\mu\alpha}g_{\nu\beta}-g_{\mu\beta}g_{\nu\alpha})
+\psi_{2}^{A}(g_{\mu\beta}v_{\nu}v^{\prime}_{\alpha}-g_{\nu\beta}v_{\mu}v^{\prime}_{\alpha}
+g_{\nu\alpha}v_{\beta}v^{\prime}_{\mu}-g_{\mu\alpha}v_{\beta}v^{\prime}_{\nu})
\nonumber\\& &{}
+\psi_{3}^{A}(g_{\alpha\mu}v_{\nu}v_{\beta}-g_{\nu\alpha}v_{\mu}v_{\beta}
+g_{\nu\beta}v^{\prime}_{\alpha}v^{\prime}_{\mu}-g_{\mu\beta}v^{\prime}_{\alpha}v^{\prime}_{\nu})
+\psi_{4}^{A}(v^{\prime}_{\mu}v_{\nu}-v^{\prime}_{\nu}v_{\mu})g_{\alpha\beta}\nonumber\\
&&{}+\psi_{5}^{A}(v^{\prime}_{\mu}v_{\nu}-v^{\prime}_{\nu}v_{\mu})v^{\prime}_{\alpha}v_{\beta},
\nonumber\\\psi_{\mu\nu}^{\alpha\beta,\,S}&=&\psi_{1}^{S}g_{\alpha\beta}g_{\mu\nu}+
\psi_{2}^{S}(g_{\mu\alpha}g_{\nu\beta}+g_{\mu\beta}g_{\nu\alpha})
+\psi_{3}^{S}(g_{\mu\beta}v_{\nu}v^{\prime}_{\alpha}+g_{\nu\beta}v_{\mu}v^{\prime}_{\alpha}
+g_{\nu\alpha}v_{\beta}v^{\prime}_{\mu}+g_{\mu\alpha}v_{\beta}v^{\prime}_{\nu})\nonumber\\
& &
{}+\psi_{4}^{S}(g_{\alpha\mu}v_{\nu}v_{\beta}+g_{\nu\alpha}v_{\mu}v_{\beta}
+g_{\nu\beta}v^{\prime}_{\alpha}v^{\prime}_{\mu}+g_{\mu\beta}v^{\prime}_{\alpha}v^{\prime}_{\nu})
+\psi_{5}^{S}(v^{\prime}-v)_{\mu}(v^{\prime}-v)_{\nu}v^{\prime}_{\alpha}v_{\beta}
\nonumber\\&&{}+\psi_{6}^{S}(v^{\prime}-v)_{\mu}(v^{\prime}-v)_{\nu}g_{\alpha\beta}
+\psi_{7}^{S}(v^{\prime}+v)_{\mu}(v^{\prime}+v)_{\nu}v^{\prime}_{\alpha}v_{\beta}
+\psi_{8}^{S}(v^{\prime}+v)_{\mu}(v^{\prime}+v)_{\nu}g_{\alpha\beta}\nonumber\\
&&{}+\psi_{9}^{S}v_{\beta}v^{\prime}_{\alpha}g_{\mu\nu},
\end{eqnarray}
introduce other universal parameters in the leading order
\begin{eqnarray}
\label{xisa} \langle \Sigma\mid\bar h_v \Gamma^{\nu}
iD_{\nu}h_{v^{\prime}}\mid\Sigma^{\prime}\rangle&=&\xi^{\alpha\beta}_{\nu}(v,v^{\prime})\bar
\Psi_{\alpha}\Gamma^{\mu\nu}\Psi_{\beta}^{\prime},\nonumber\\
\langle \Sigma\mid\bar h_v \Gamma^{\nu} (-i\overleftarrow
D_{\nu})h_{v^{\prime}}\mid\Sigma^{\prime}\rangle&=&\bar\xi^{\beta\alpha}_{\nu}(v^{\prime},v)\bar
\Psi_{\alpha}\Gamma^{\mu\nu}\Psi_{\beta}^{\prime},
\end{eqnarray}
as usual, $\xi^{\alpha\beta}_{\nu}(v,v^{\prime})$ can be
decomposed into the general form
\begin{eqnarray}
\label{dxisa}
\xi^{\alpha\beta}_{\nu}(v,v^{\prime}&=&\xi_1(v+v^{\prime})_{\nu}g_{\alpha\beta}
+\xi_2(v^{\prime}-v)_{\nu}g_{\alpha\beta}+\xi_3(v+v^{\prime})_{\nu}v_{\beta}v^{\prime}_{\alpha}
\nonumber\\&&{}
+\xi_4(v^{\prime}-v)_{\nu}v_{\beta}v^{\prime}_{\alpha}
+\xi_5v^{\prime}_{\alpha}g_{\beta\nu}+\xi_6v_{\beta}g_{\alpha\nu},
\end{eqnarray}
the equation of motion implies that
$v^{\prime\nu}\xi^{\alpha\beta}_{\nu}=0$ and
$v^{\prime\nu}\psi_{\mu\nu}^{\alpha\beta}=0$ from which we can
derive relations
\begin{eqnarray}
\label{presa}
w\psi_{3}^{A}-\psi_{2}^{A}+\psi_{3}^{S}+\psi_{4}^{S}&=&0,\nonumber\\
w\psi_{2}^{A}-\psi_{1}^{A}-\psi_{3}^{A}+\psi_{2}^{S}+w\psi_{3}^{S}+\psi_{4}^{S}&=&0,\nonumber\\
w\psi_{4}^{A}+\psi_{1}^{S}+(1-w)\psi_{6}^{S}+(1+w)\psi_{8}^{S}&=&0,\nonumber\\
\psi_{2}^{A}+w\psi_{5}^{A}+\psi_{3}^{s}+(1-w)\psi_{5}^{S}+(1+w)\psi_{7}^{S}+\psi_{9}^{S}&=&0,\nonumber\\
(1+w)\psi_{8}^{S}-\psi_{4}^{A}-(1-w)\psi_{6}^{S}&=&0,\nonumber\\
\psi_{4}^{S}+(w-1)\psi_{5}^{S}-\psi_{3}^{A}-\psi_{5}^{A}+(1+w)\psi_{7}^{S}&=&0,\nonumber\\
\end{eqnarray}
and
\begin{eqnarray}
\label{xres} (1+w)\xi_1+(1-w)\xi_2=0,\nonumber\\
(1+w)\xi_3+(1-w)\xi_4+\xi_6=0,
\end{eqnarray}
take the difference of the two terms in Eqs.~(\ref{xisa}) and use
(\ref{deria}) we can reach
\begin{eqnarray}
\xi_1&=&\frac{w-1}{w+1}\frac{c_1}{2}\bar\Lambda,\nonumber\\
\xi_3&=&\frac{w-1}{w+1}\frac{c_2}{2}\bar\Lambda-\xi_6,\nonumber\\
\xi_2&=&\frac{c_1}{2}\bar\Lambda,\nonumber\\
\xi_4&=&\frac{c_2}{2}\bar\Lambda,\nonumber\\ \xi_5&=&\xi_6,
\end{eqnarray}
where $c_1, c_2$ parameterize the matrix element
\begin{equation}
\langle \Sigma\mid\bar h_v \Gamma
h_{v^{\prime}}\mid\Sigma^{\prime}\rangle=(c_1g_{\alpha\beta}+c_2v_{\beta}v_{\alpha}^{\prime})
\bar\Psi^{\alpha}\Gamma \Psi^{\prime \beta}.
\end{equation}
The matrix elements for the kinetic energy and chromo-magnetic
interaction are defined similar to those for the $\Lambda_Q$
baryon
\begin{equation}
\label{kvsa} \langle \Sigma\mid\bar h_v (iD^{\perp})^2\Gamma
h_{v^{\prime}}\mid\Sigma^{\prime}\rangle=(\phi_0g_{\alpha\beta}+\bar\phi_0v_{\beta}v^{\prime}
_{\alpha})  \bar\Psi^{\alpha}\Gamma \Psi^{\prime\beta},
\end{equation}
\begin{equation}
\label{svsa} \langle \Sigma\mid\bar h_v \sigma_{\mu\nu}iG^{\mu\nu}
h_{v^{\prime}}\mid\Sigma^{\prime}\rangle=\phi_{\mu\nu}^{\alpha\beta}
 \bar\Psi_{\alpha}\sigma^{\mu\nu} \Psi^{\prime}_{\beta} ,
\end{equation}
where $\phi_{\mu\nu}^{\alpha\beta}$ bear the same decomposition as
$\psi_{\mu\nu}^{\alpha\beta}$ and they have simple relations
between each other
\begin{eqnarray}
\phi_1&=&\psi_{1}^{A},\nonumber\\
\phi_2&=&\psi_{2}^{A}-\xi_6\bar\Lambda,\nonumber\\
\phi_3&=&\psi_{3}^{A}-\xi_6\bar\Lambda,\nonumber\\
\phi_4&=&\psi_{4}^{A}-2\xi_1\bar\Lambda,\nonumber\\
\phi_5&=&\psi_{5}^{A}-2\xi_3\bar\Lambda,\nonumber\\
\phi_0&=&2\psi_{1}^{s}+\psi_{2}^{s}+(1-w)\psi_{6}^{s}+(1+w)\psi_{8}^{s}+2(1-w)\xi_2\bar\Lambda,\nonumber\\
\bar\phi_0&=&2\psi_{3}^{s}+(1-w)\psi_{5}^{s}+(1+w)\psi_{7}^{s}+2(1-w)\xi_4\bar\Lambda,
\end{eqnarray}
the normalization condition is that $\phi_{1}(1)=A\lambda_2,
\,\phi_{0}(1)=B\lambda_1$ where $A$ is $-1/2, 1/2$ and $B$ is $1,
-1$ for $\Sigma_{Q}^{*}, \Sigma_{Q}$ respectively. At zero recoil
$\lambda_1$ can be expressed via $\phi_0$
\begin{equation}
\phi_{0}(1)=\phi_{1}(1)-2[\phi_{2}(1)-\phi_{3}(1)]-3\phi_{4}(1)\;.
\end{equation}
\end{appendix}

%%%%%%%%%%%%%%%%%%%%%%%%%%%%%%%%%%%%%%%%%%%

%%%%%%%%%%%%%%%%%%%%%%%%%%%%%%%%%%%%%%%%%%%%%%%%%%%%%%%%%%%%%%%%%%%%%%%
\newpage
{\bf Figure Captions}
\begin{center}
\begin{minipage}{12cm}
{\sf Fig. 1.}{\quad Non-vanishing diagrams for the kinetic energy
: (a) perturbative contribution, (b) to (e) gluon-condensate. The
kinetic energy operator is denoted by a white square, the
interpolating baryon currents by black circles. Heavy-quark
propagators are drawn as double lines. Diagrams (b) to (e) are
calculated in Fock-Schwinger gauge. The lower right vertices of
those diagrams are set to the origin in coordinate space.}
\end{minipage}
\end{center}

\begin{center}
\begin{minipage}{12cm}

{\sf Fig. 2.}{\quad Non-vanishing diagrams for the chromo-magnetic
interaction: (a) perturbative contribution, (b) gluon-condensate,
(c) quark-condensate. The chromo-magnetic interaction
(velocity-changing current) operator is denoted by a white square,
the interpolating baryon currents by black circles.}
\end{minipage}
\end{center}

\begin{center}
\begin{minipage}{120mm}
{\sf Fig. 3.}{\quad Sum rules for $\Lambda_Q$ baryon: (a) for
$j^{v}_{1}$, (b) for $j^{v}_{2}$. The dash-dotted, dashed and
solid curves correspond to the threshold $\omega_c=2.4, 2.6,
2.8\;\mbox{GeV}$, respectively. The working region is
$\mbox{T}=0.8-1.2\,\mbox{GeV}$.}
\end{minipage}
\end{center}

\begin{center}
\begin{minipage}{120mm}
{\sf Fig. 4.}{\quad Sum rules of the kinetic energy for $\Sigma_Q$
baryon: (a) for $j^{v}_{1}$, (b) for $j^{v}_{2}$. The dash-dotted,
dashed and solid curves correspond to the threshold $\omega_c=2.9,
3.1, 3.3\;\mbox{GeV}$, respectively. The working region is
$\mbox{T}=0.7-1.1\,\mbox{GeV}$.}
\end{minipage}
\end{center}

\begin{center}
\begin{minipage}{12cm}
{\sf Fig. 5.}{\quad Covariant sum rules of the kinetic energy: (a)
for $\Lambda_Q$ baryon, (b) for $\Sigma_Q$ baryon. The
dash-dotted, dashed and solid curves correspond to $\omega_c=2.4,
2.6, 2.8\;\mbox{GeV}$ for $\Lambda_Q$ baryon, and $\omega_c=2.9,
3.1, 3.3\;\mbox{GeV}$ for $\Sigma_Q$ baryon. The working region is
$\mbox{T}=0.8-1.2\,\mbox{GeV}$ for $\Lambda_Q$ baryon and
$\mbox{T}=0.7-1.1\,\mbox{GeV}$ for $\Sigma_Q$ baryon.}
\end{minipage}
\end{center}

\begin{center}
\begin{minipage}{12cm}
{\sf Fig. 6.}{\quad Sum rules for the chromo-magnetic interaction.
The dash-dotted, dashed and solid curves correspond to
$\omega_c=2.9, 3.1, 3.3\;\mbox{GeV}$. The working region is
$\mbox{T}=0.7-1.1\,\mbox{GeV}$.}
\end{minipage}
\end{center}
\newpage
\vspace{7cm} %\vfill\bigskipamount
\begin{figure}
\vfill \epsfxsize=12cm
\centerline{\epsffile{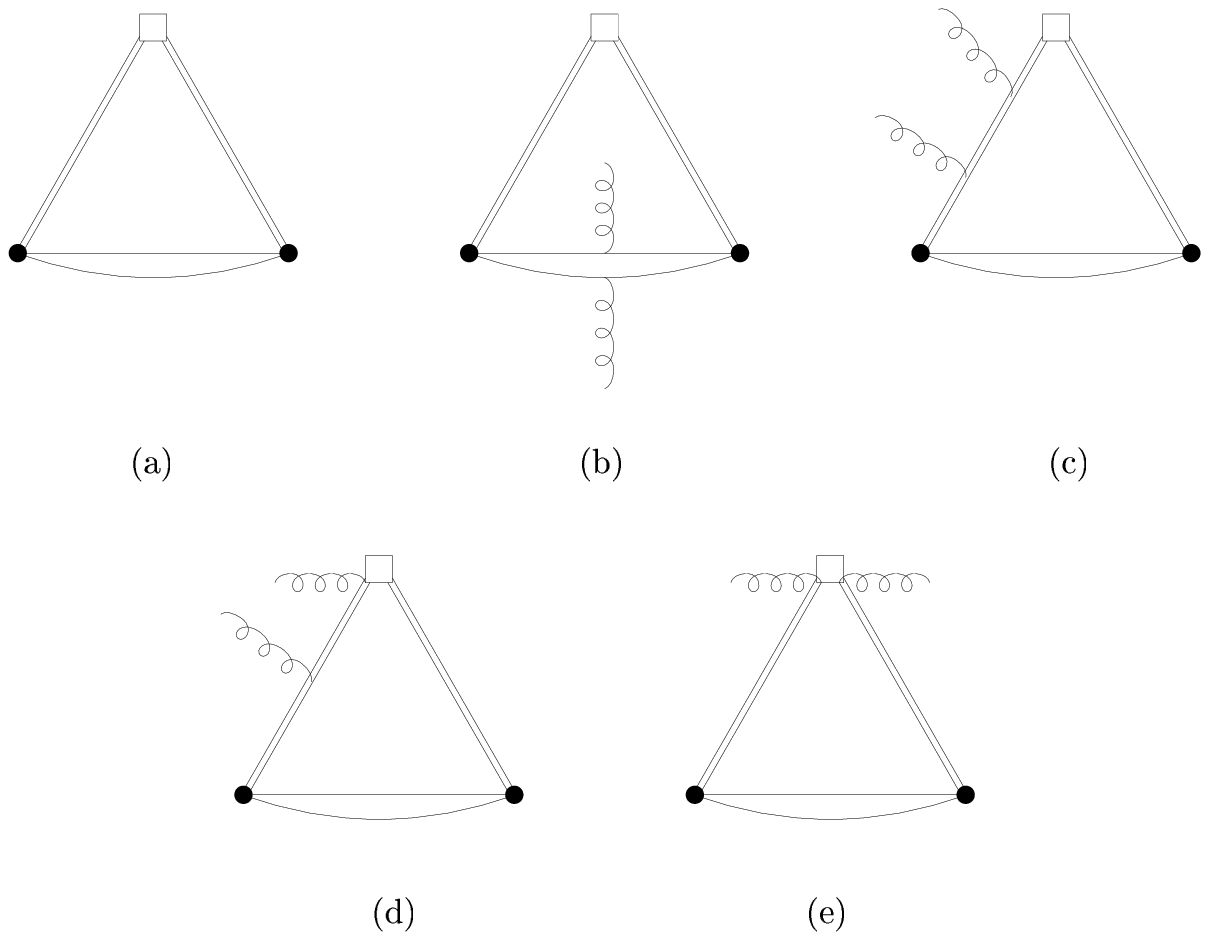}}\mbox{}\vfill \center{\mbox{Fig.
1.}}
\end{figure}

\begin{figure}\vfill
 \epsfxsize=12cm \centerline{\epsffile{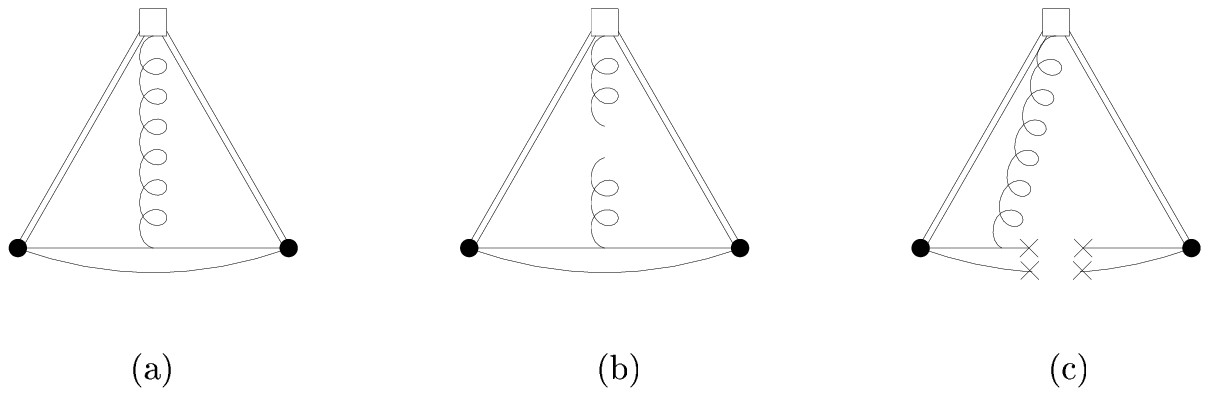}}
\mbox{}\vfill\center{\mbox{Fig. 2.}}
\end{figure}
\newpage
\begin{minipage}{15cm}
\begin{minipage}{7cm}
\begin{figure}
\epsfxsize=7cm \centerline{\epsffile{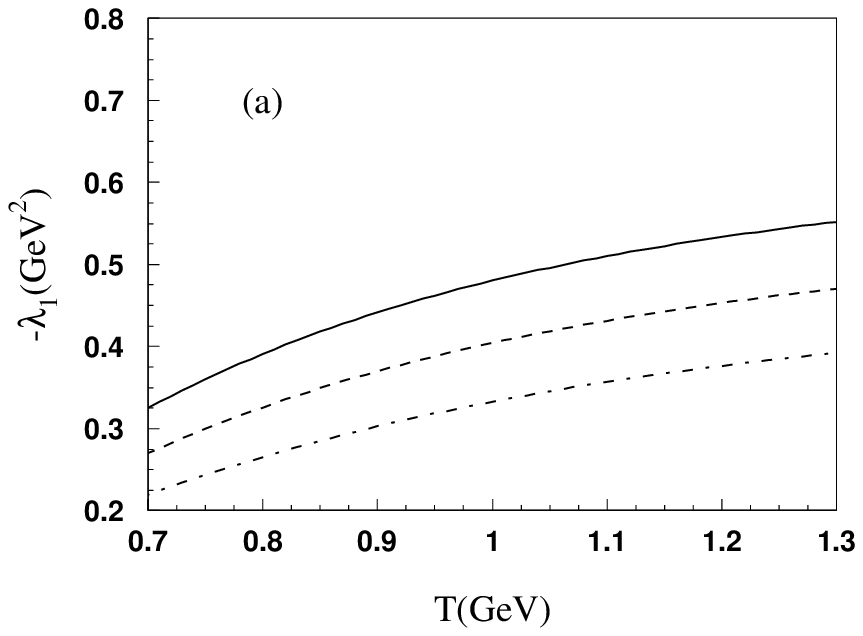}}
\end{figure}
\mbox{}
\end{minipage}
\begin{minipage}{7cm}
\begin{figure}
\epsfxsize=7cm \centerline{\epsffile{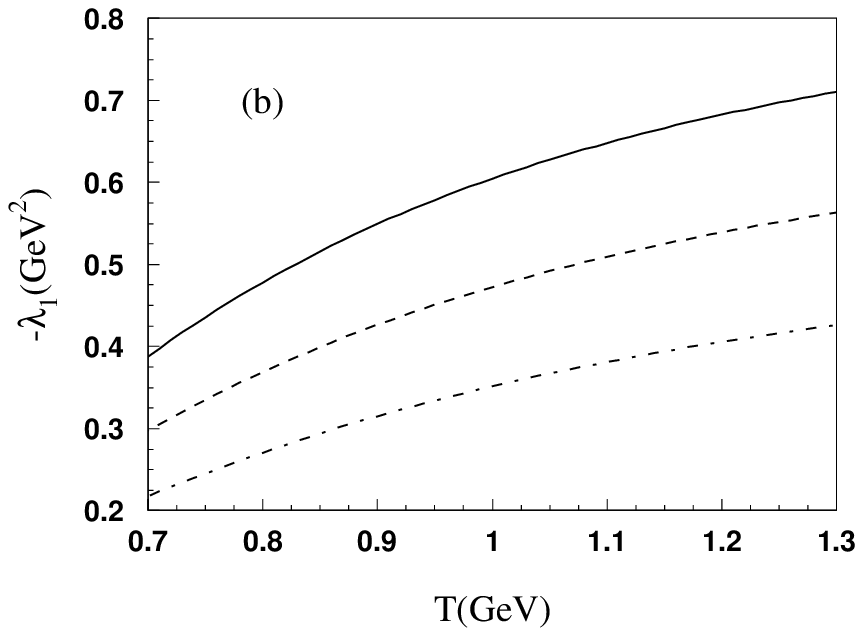}}
\end{figure}
\mbox{}
\end{minipage}
\center{\mbox{Fig. 3.}}% \vfill \mbox{}
\end{minipage}
\vfill
\begin{minipage}{15cm}
\begin{minipage}{7cm}
\begin{figure}%\vfill
\epsfxsize=7cm \centerline{\epsffile{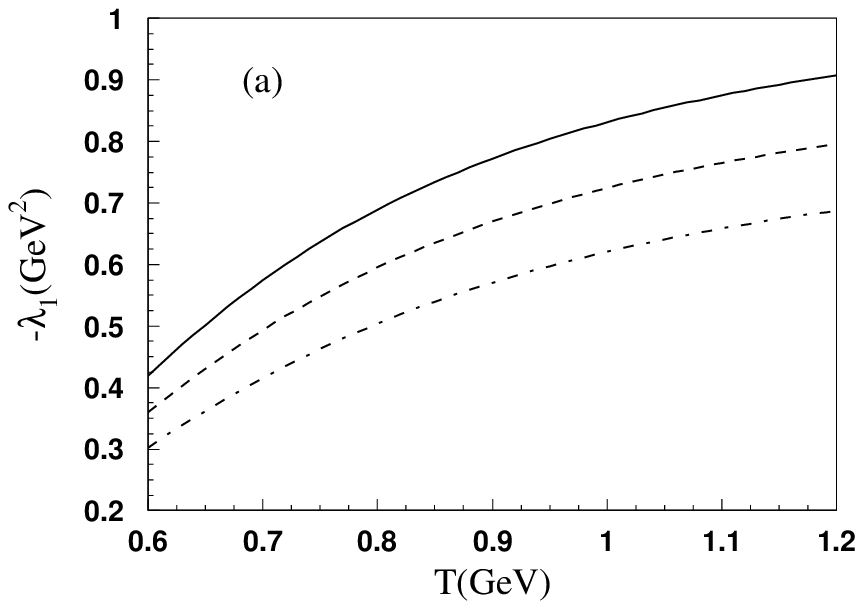}}
\end{figure}\mbox{}
\end{minipage}
\begin{minipage}{7cm}
\begin{figure}
\epsfxsize=7cm \centerline{\epsffile{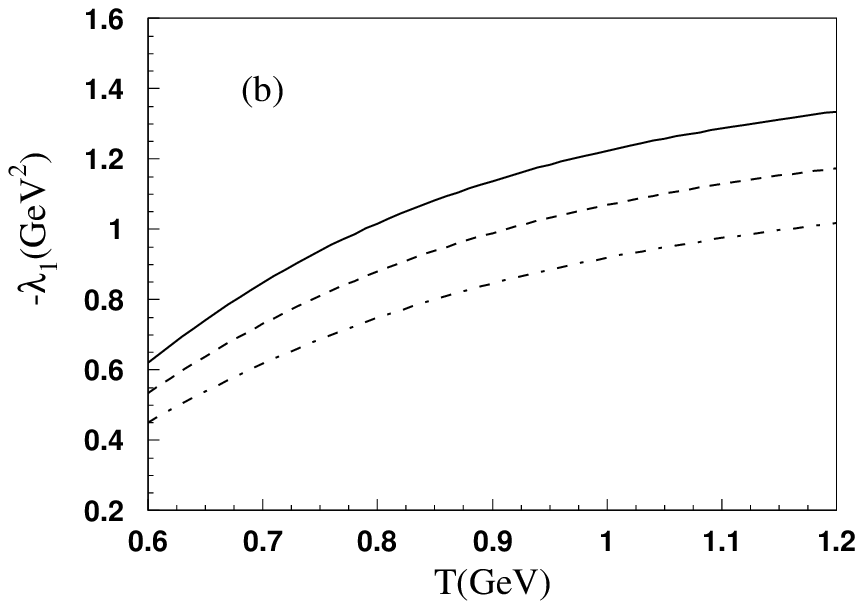}}
\end{figure}
\mbox{}
\end{minipage}
\center{\mbox{Fig. 4.}}

%\vfill\mbox{}
\end{minipage}
\vfill
\begin{minipage}{15cm}
\begin{minipage}{7cm}

\begin{figure}
\epsfxsize=7cm \centerline{\epsffile{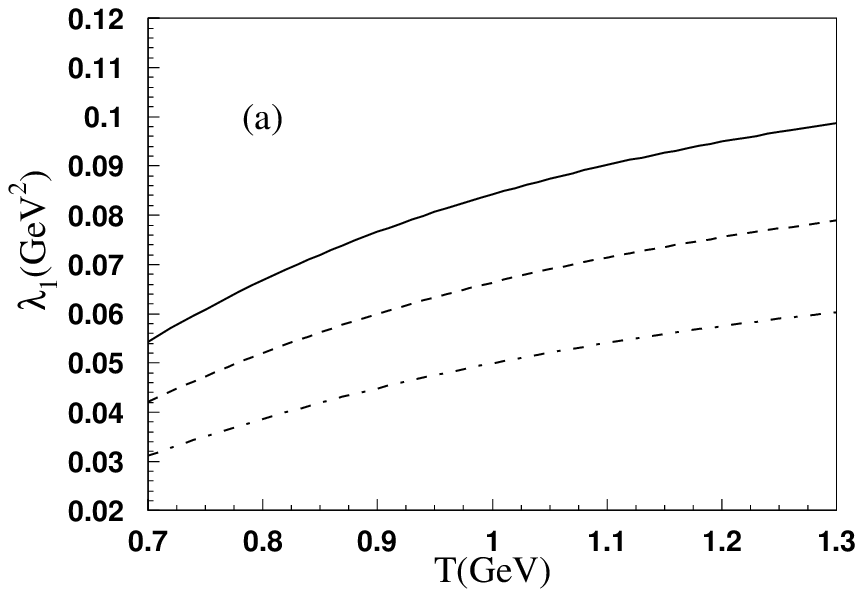}}
\end{figure}
\mbox{}
\end{minipage}
%\hfill
\begin{minipage}{7cm}
\begin{figure}
\epsfxsize=7cm \centerline{\epsffile{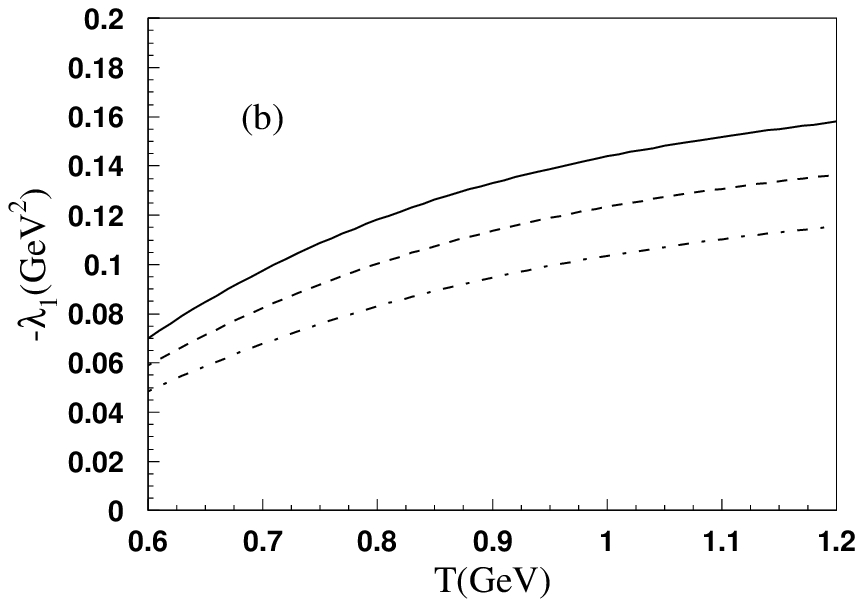}}
\end{figure}
\mbox{}
\end{minipage}
\center{\mbox{Fig. 5.}}%\mbox{}
\end{minipage}

\begin{minipage}{15cm}
\begin{figure}
\epsfxsize=7cm \centerline{\epsffile{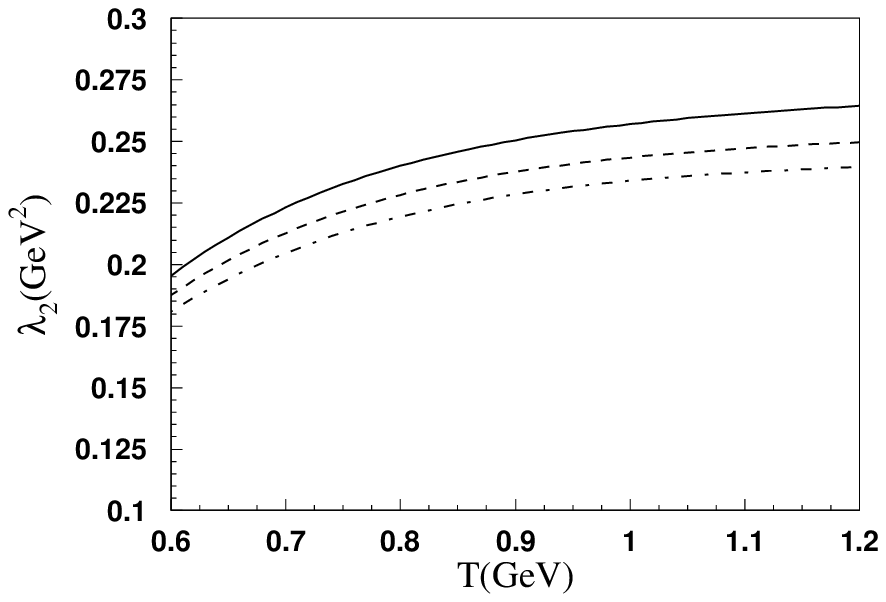}}
\end{figure}
\center{\mbox{Fig. 6.}}
\end{minipage}

\end{document}